\documentclass[preprint,authoryear,12pt]{elsarticle}

\usepackage{epsfig}

\usepackage{amssymb}

\usepackage[ps2pdf,%
a4paper=true,%
breaklinks=true,%
colorlinks=true,%
pdfauthor={First Author et al.},%
pdftitle={Template for manuscripts in Advances in Space Research}%
]{hyperref}

\journal{Advances in Space Research}

\begin{document}

\begin{frontmatter}

\title{Eruption of the magnetic flux rope in a fast decayed active region}

\author{Shangbin Yang\corref{cor}}
\address{Key Laboratory of Solar Activity, National Astronomical
Observatories, Chinese Academy of Sciences, Beijing 100012, China}
\cortext[cor]{Corresponding author} \ead{yangshb@bao.ac.cn}
\author{Wenbin Xie}
\address{Jilin Normal University, 136000 Siping, Jilin
Province, China}
\author{Jihong Liu}
\address{Shijiazhuang University, 050035 Shijiazhuang, Hebei Province, China}

\begin{abstract}
An isolated and fast decayed active region (NOAA 9729) was observed
when passing through solar disk. There is only one CME related with
it that give us a good opportunity to investigate the whole process
of the CME. Filament in this active region rises up rapidly and then
hesitates and disintegrates into flare loops. The rising filament
from EIT images separates into two parts just before eruption. A new
filament reforms several hours later after CME and the axis of this
new one rotates clockwise about $22^{\circ}$ comparing with that of
the former one. We also observed a bright transient J-shaped X-ray
sigmoid immediately appears after filament eruption. It quickly
develops into a soft X-ray cusp and rises up firstly then drops
down. Two magnetic cancelation regions have been observed clearly
just before filament eruption. Moreover, the magnetic flux rope
erupted as the magnetic helicity approach the maximum and the
normalized helicity is -0.036 when the magnetic flux rope erupted,
which is close to the prediction value of \citet{Zhang08} based on
the theoretical non-linear force-free model.
\end{abstract}


\end{frontmatter}

\parindent=0.5 cm

\section{Introduction}
\label{sec:Introduction}
 Coronal mass ejections (CMEs) have been one
of the outstanding problems of solar physics. The nature of driver
and initiation mechanism for the sudden explosive release of the
stored free magnetic energy are still unclear. It is proposed that
CMEs are that metastable magnetic configurations are followed by
some finite perturbation or some additional energy build-up and make
a eventual catastrophic transition to a lower energy state or to
non-equilibrium state \citep[Ref.][]{PF02,Lin03}. There are already
several mechanisms to cause the catastrophic transition have been
proposed such as slow reduction of the overlying flux
\citep{FP95,PF90}, photospheric converging and shear motions
\citep{Fob94,Ant94}, flux emergence \citep{Fey95}, etc.

As a candidate for the metastable configuration, \citet{Stu01}
considered a long twisted flux tube, anchored at both ends in the
photosphere with overlying magnetic arcade. They argue from a simple
order-of-magnitude calculation that part of the flux tube will open
up to infinity if there are 1-2 full winds for each field line in
the flux tube even this configuration is stable according to linear
MHD stability theory. \citet{Fan05} carried out simulations in a
spherical geometry of the evolution of coronal magnetic field as an
a twisted magnetic flux rope emerges slowly into a preexisting
coronal potential arcade field. She find that the flux tube becomes
kinked and ruptures through the arcade field and cause a eruption
when the twist in the emerged tube reaches a critical amount.
\citet{TK05} use the flux rope model of \citet{TD99} as the initial
condition to get a simulation which have a good agreement with the
development of helical shape and the rise profile of a failed
filament eruption described by \citet{Ji03}. Kinking movement in the
eruption of filament are also usually observed \citep{Liu07,Liu08}.
In addition, soft X-ray images of solar active regions frequently
show S- or inverse-S (sigmoidal) morphology. \citet{Can99} found
that active regions containing X-ray sigmoids are more likely to
erupt and many eruption also associated with sigmoid structure.
While the question of whether there exit highly twisted flux ropes
with more than one wind between anchored ends susceptible to the
kink instability as precursors for eruptive flux tube remains a
topic of debate \citep[e.g.][]{RK96,Lea03,Lek05,RL05} when these
authors investigate the relation between the shape of X-ray sigmoid
and eruption.

\citet{GF06} demonstrated the partial expulsion of a three
dimensional magnetic flux rope erupts when enough twist has emerged
to induce a loss of equilibrium. After multiple reconnections at
current sheets that form during the eruption, the rope breaks in
two, so that only a part of it escapes. The "degree of emergence" of
a pre-eruption flux rope, whether it possess bald-patch (BP) or
whether it is high enough in the coronal to possess an X-line
determines whether the rope is expelled totally or partially. Their
simulation result is well consistent with the partial eruption model
described by \citet{Gil01}. But should a twisted or kinked or
magnetic flux tube need kink instability or twist over such
threshold to eruption? Is it kink instability or kink-induced
instability in a filament eruption \citep{Gil07}? For example, the
kink instability may be occurring in conjunction with a breakout
scenario \citep{Wil05}. \citet{Low01} also argued that a force-free
magnetic field in the unbounded space outside a sphere cannot be in
equilibrium if the amount of detached flux is too large compared to
the amount of anchored flux. Eruption of magnetic flux tube can also
be caused by another instabilities such as torus instability
\citep{KT06} and Ballooning instability \citep[e.g.][]{Fon01}. So it
is essential to investigate the whole eruptive process of a magnetic
flux tube since emergence and evolution of structure in different
wavelength especially the soft X-ray sigmoid for study how the
metastable state is established and how to loss equilibrium and
erupt for this magnetic flux tube.

On the other hand, \citet{Zhang06} pointed out that the accumulation
of magnetic helicity in the corona plays a significant role in
storing magnetic energy. They propose a conjecture that there is an
upper bound on the total magnetic helicity that a force-free field
can contain. The accumulation of magnetic helicity in excess of this
upper bound would initiate a non-equilibrium situation, resulting in
a CME expulsion as a natural product of coronal evolution.
\citet{Ber84} proposed the concept of relative magnetic helicity and
the formula to get the accumulated magnetic helicity across a
surface into a volume from the movement in that surface.
\citet{Cha01} firstly calculated the accumulated magnetic helicity
by applying LCT (Local Correlation Tracking) to MDI data. However,
the relation between accumulated magnetic helicity and eruption of
magnetic flux tube has not been checked in the past work. What is
the accumulated helicity for a emerging flux tube when it erupts? We
need to investigate the process of magnetic helicity accumulation in
an eruption.

NOAA 9729 is an active region emerging on Dec. 05, 2001. Its initial
tilt angle is almost perpendicular to the equator of sun and then
develop to a bipolar active region. It rapidly dispersed in the
following days and disappeared on Dec. 08, 2001 from white light
(WL) image. The evolution of such photospheric concentrations are
usually been explained in terms of the rising of very distorted flux
tubes. \citet{wea70,wea72} also noticed the almost random
distribution of the starting tilt of emerging bipoles, which
subsequently became more parallel to the equator, and proposed that
this was caused by the emergence of twisted flux tubes. Especially,
a kinked flux tube arches upward and evolves into a buckled loop
with a local change of tube orientation at the loop apex that
exceeds 90 degrees from the original direction of the tube
\citet{Fan99}. The characteristic of this emerging active region
implies that there are strong twist in it. Moreover, there is only
one CME related with it since emergence. This give us a good
opportunity to investigate the whole process of a CME for a strong
twisted flux tube.

In this paper we use multiple wavelengths and instruments to
investigate the whole CME process.  We also calculate the
accumulated magnetic helicity since this active region emerged. We
give a quick survey of the instrument and data sets used in this
study (Sec.~\ref{sec:data}). We describe the evolution of white
light and emerging speed in Sec. ~\ref{sec:wl}. we describe the
evolution of H-alpha, EIT, X-ray, and corresponding CMEs from
Sec.~\ref{sec:halpha} to \ref{sec:CME}. We calculate accumulated
magnetic helicity in Sec.~\ref{sec:helicity}. The summary and
discussion is represented in Sec.~\ref{sec:summary}.
\section{OBSERVATIONS AND DATA ANALYSIS}
\label{sec:observation}
\subsection{Instrumentation and Data}
\label{sec:data} We use data of SOHO/MDI to investigate the
evolution of White light image and line-of-sight magnetograms
\citep{Sche95}. We use data of high-resolution global $H\alpha$
network to investigate the evolution of Chromosphere. We use data of
Extreme Ultraviolet Imaging Telescope to investigate the evolution
of low Corona \citep[EIT;][]{Del95}. We use soft X-ray images
obtained with the Soft X-ray Telescope (SXT) on board the Yohkoh
satellite to investigate the evolution of high Corona \citep{Tsu91}.
We use Large Angle Spectrometric Coronagraph \citep[LASCO;][]{Bru95}
aboard SOHO to investigate the CME associated with this active
region.
\subsection{Estimation of emerging speed}
\label{sec:wl} 
\begin{figure}[htb]
 \centering
 \includegraphics[angle=0,scale=0.6]{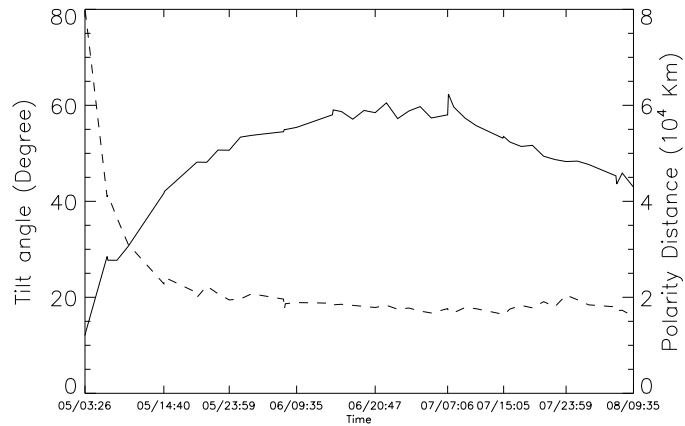}
 \caption{Evolution of tilt angle (dashed line) and distance (solid line) between two polarities.}
 \label{fig:DT}
 \end{figure}

NOAA 09729 emerged from the convective zone at about
03:26UT on December 05, 2001 seen from WL image. It developed to a
bipolar active region rapidly and disappeared on Dec. 08, 2001. The
evolution of tilt angle and the distance between two polarities
considering the barycenters of the leading and following polarities
are shown in Fig.~\ref{fig:DT}. Its initial title angle  is almost
perpendicular to the equator of sun and still disobeyed Joy's law
when this active region disappeared despite the connecting line
between two polarities became more parallel to the equator. The
maximum distance ($dmax$) between two polarities is at 07:06UT on
December 7. If we suppose the coronal part of one active region is
represented by a single semicircular loop. The initial distance
($dmin$) will be the chord of the semicircle and the maximum
distance ($dmax$) will be the diameter of this semicircle. The
emerging hight in this time interval is approximately $\sqrt{(dmax
^2-dmin ^2)/4}=30.6Mm$ under this assumption and the corresponding
average emerging speed is about 0.164km/s. The mean magnetic field
in this active region is about 100 Gauss and the corresponding
${\textrm{Alfv}\acute{\textrm{e}}\textrm{n}}$ speed at photosphere
is about 8.9Km/s . The emerging speed will be about 0.018 of local
$V_A$. Fig.~\ref{fig:mdi_WL} shows the time sequence of WL images.

\subsection{$\textrm{H}{\alpha}$ Evolution}
\label{sec:halpha}
 Fig.~\ref{fig:halpha} shows the evolution of NOAA
9729 in $\textrm{H}{\alpha}$. Top two rows
(Fig.~\ref{fig:halpha}a-f) show the evolution of
$\textrm{H}{\alpha}$ from December 05 to 09, 2001. The filament in
this active region didn't appeared before 22:30 UT December 4. At
16:59UT on December 5, a inverse-S shaped filament has appeared in
the $\textrm{H}{\alpha}$ locating above polarity inversion line
(PIL) which is shown in Fig.~\ref{fig:halpha}b. The shape of this
filament didn't changed a lot after this moment while the middle
channel of this filament labeled by red solid line in
Fig.~\ref{fig:halpha}c separated into several parts and is not clear
than the initial one on December 5. The filament still existed after
eruption at about 02:34UT on December 7 and there is an obvious
clockwise rotation of the middle channel of the filament. The
clockwise rotation angle of the filament middle channel is
approximately $22^{\circ}$ denoted in the Fig.~\ref{fig:halpha}. The
filament almost kept the same shape in the next two days after this
active region disappeared from WL, which can be seen in the
Fig.~\ref{fig:halpha} (e-f).

Bottom two rows (Fig.~\ref{fig:halpha}g-n) show the evolution detail
in $\textrm{H}{\alpha}$ when eruption occurred on December 07, 2001.
There are two stages for the filament eruption. Firstly, the left
part in the middle filament channel became thin and disappeared
gradually from Fig.~\ref{fig:halpha}g-j. The left part of the middle
filament channel has disappeared entirely at 01:45UT. The right part
of the filament disappeared totally before 02:12UT in
Fig.~\ref{fig:halpha}. One interesting phenomena is that the right
part of filament became thicker before eruption which is different
with the left one. At 03:41UT after eruption, two bright foot-points
regions have appeared and the brightness decreased with time. The
shape of the two bright regions is consistent with cancelation
region of magnetic field region in Fig.~\ref{fig:mdi}f. The filament
above the polarity inversion line (PIL) also gradually appeared
again after the eruption, which is shown from
Fig.~\ref{fig:halpha}l-n.
\subsection{EIT evolution}
\label{sec:EIT} Fig. 4 shows the detail evolution of NOAA 9729 EIT
images in the 171 {\AA} filter when the eruption occurred on
December 7. The filament in EUV wavelength can be seen clearly from
the Fig.~\ref{fig:EIT}. There are similar stages for the filament
eruption from EUV comparing with that in $\textrm{H}{\alpha}$. Two
square boxes in Fig. ~\ref{fig:EIT}a denotes two EUV bright points.
The left part risen up labeled by red arrow in Fig.~\ref{fig:EIT}b
and it did not erupt immediately. In the same time, The upper EUV
bright point disappeared . The right part of the filament risen up
subsequently, which is labeled by blue arrows in
Fig.~\ref{fig:EIT}b-c. In Fig.~\ref{fig:EIT}d, the risen filament
separated in to two parts that labeled by two red arrows. At 02:34UT
on December 07, the filament eruption occurred. Fig.~\ref{fig:EIT}e
shows the rising filament and a kinked-like structure can also been
found. After eruption, two bright EUV regions formed on the two
foot-points of this flux tube which is labeled by red boxes in
Fig.~\ref{fig:EIT}f. At 03:10UT, a bright region denoted by red
boxes above the inversion line also has formed subsequently and it
expanded along the inversion line to the opposite directions and
post-flare loops formed in the same time, which is shown from
Fig.~\ref{fig:EIT}h-j. In the following several hours, the formed
post-flare loop continued rising up and it became cooling gradually
and disappeared at last from Fig.~\ref{fig:EIT}k-l.
\subsection{Yohkoh X-ray Evolution}
\label{sec:xray}
\begin{figure*}[htb]
\centering
\includegraphics[angle=0,scale=1.0]{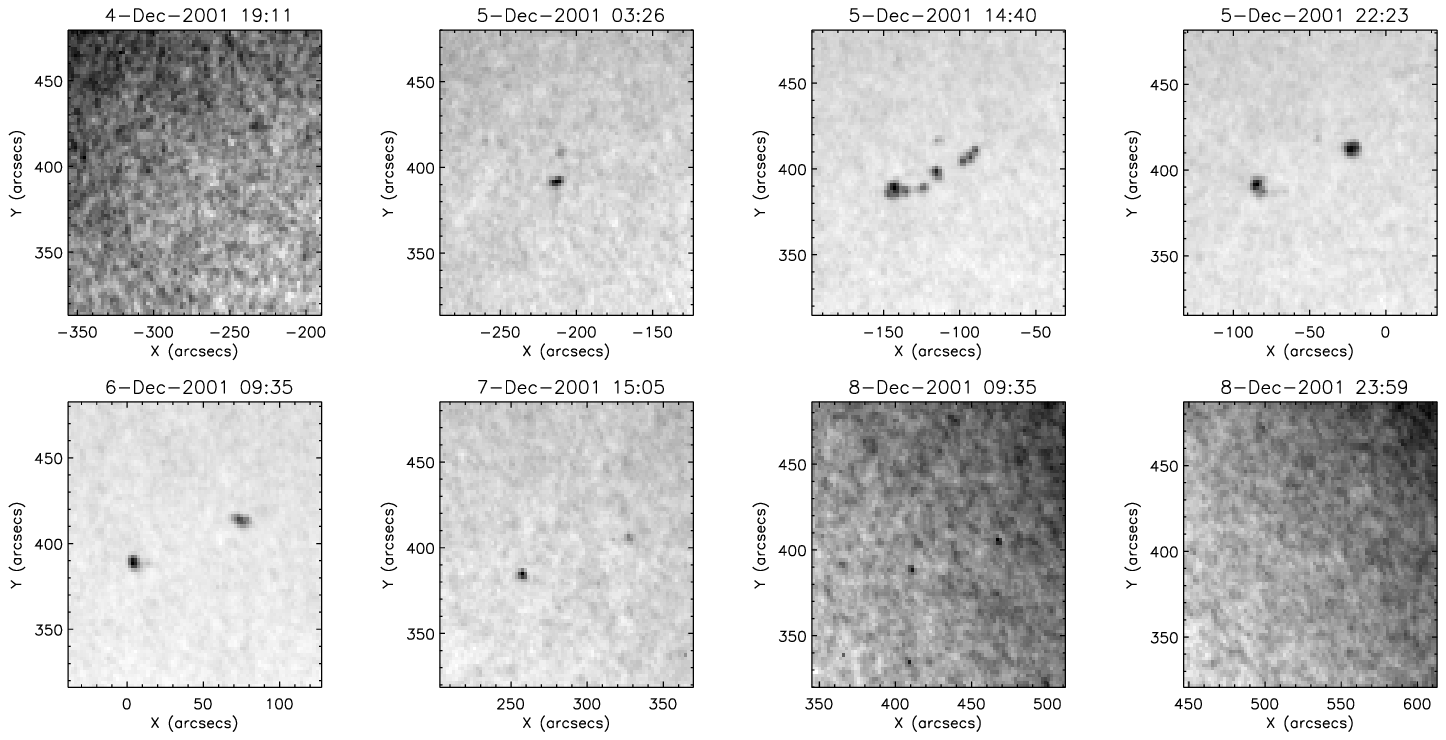}
\caption{WL evolution.} \label{fig:mdi_WL}
\end{figure*}

\begin{figure*}[htb]
\centering
\includegraphics[angle=0,scale=1.0]{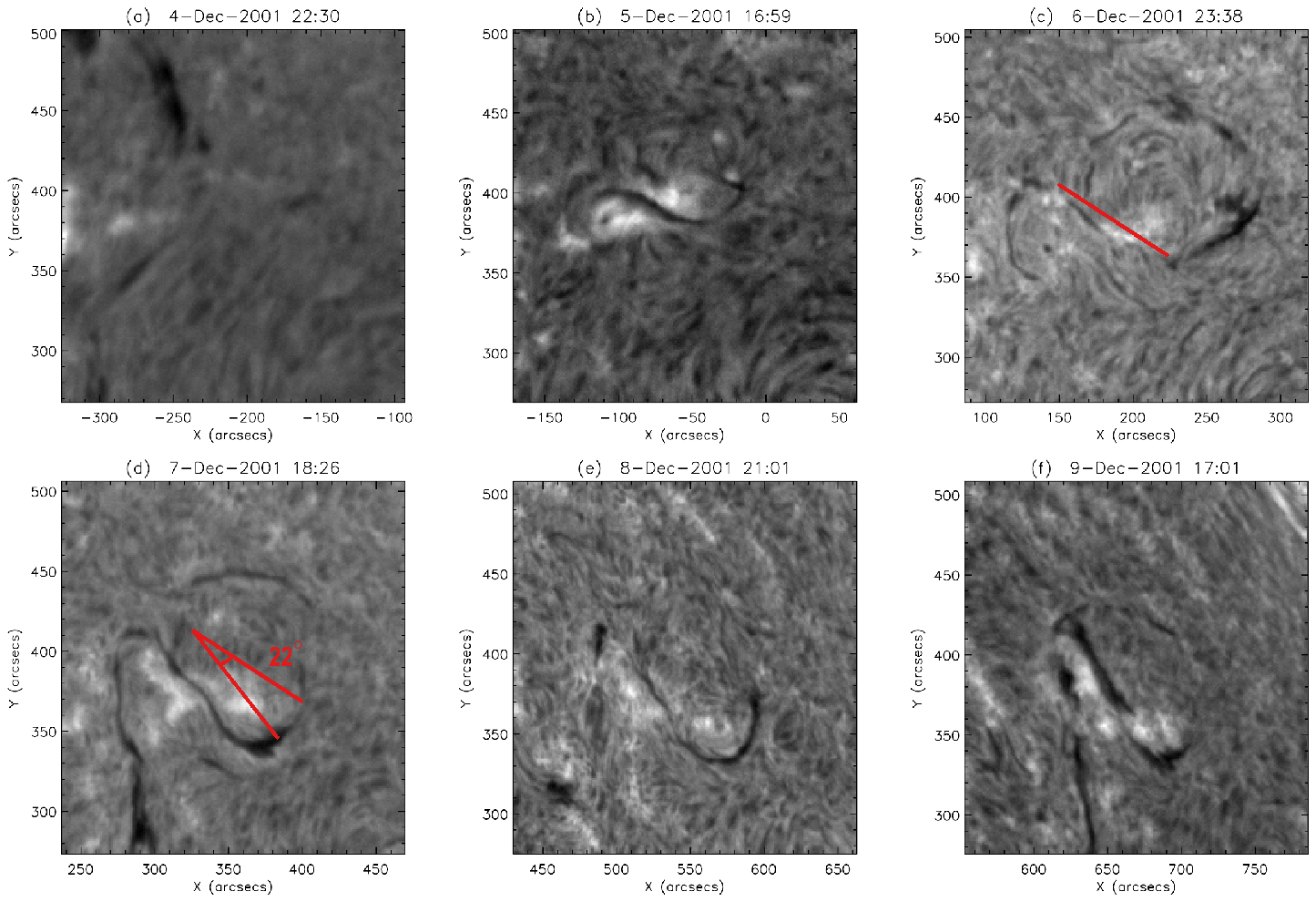}
\includegraphics[angle=0,scale=1.0]{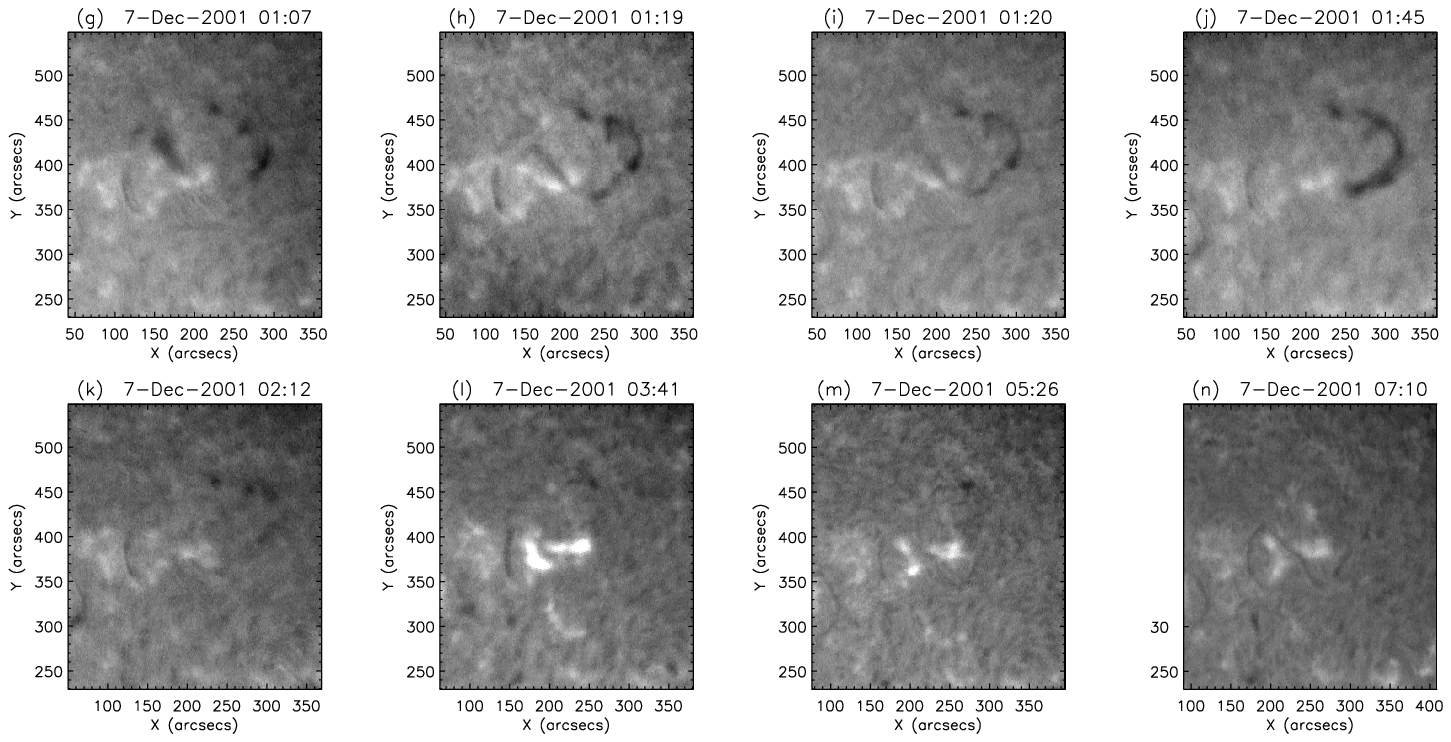}
\caption{ Time Sequence of  $\textrm{H}{\alpha}$ images. Top two
rows show the evolution of $\textrm{H}{\alpha}$ from December 04 to
09, 2001. Bottom two rows show the evolution of $\textrm{H}{\alpha}$
in the filament eruption on December 07.}
 \label{fig:halpha}
\end{figure*}

\begin{figure*}[htb]
\centering
\includegraphics[angle=0,scale=0.8]{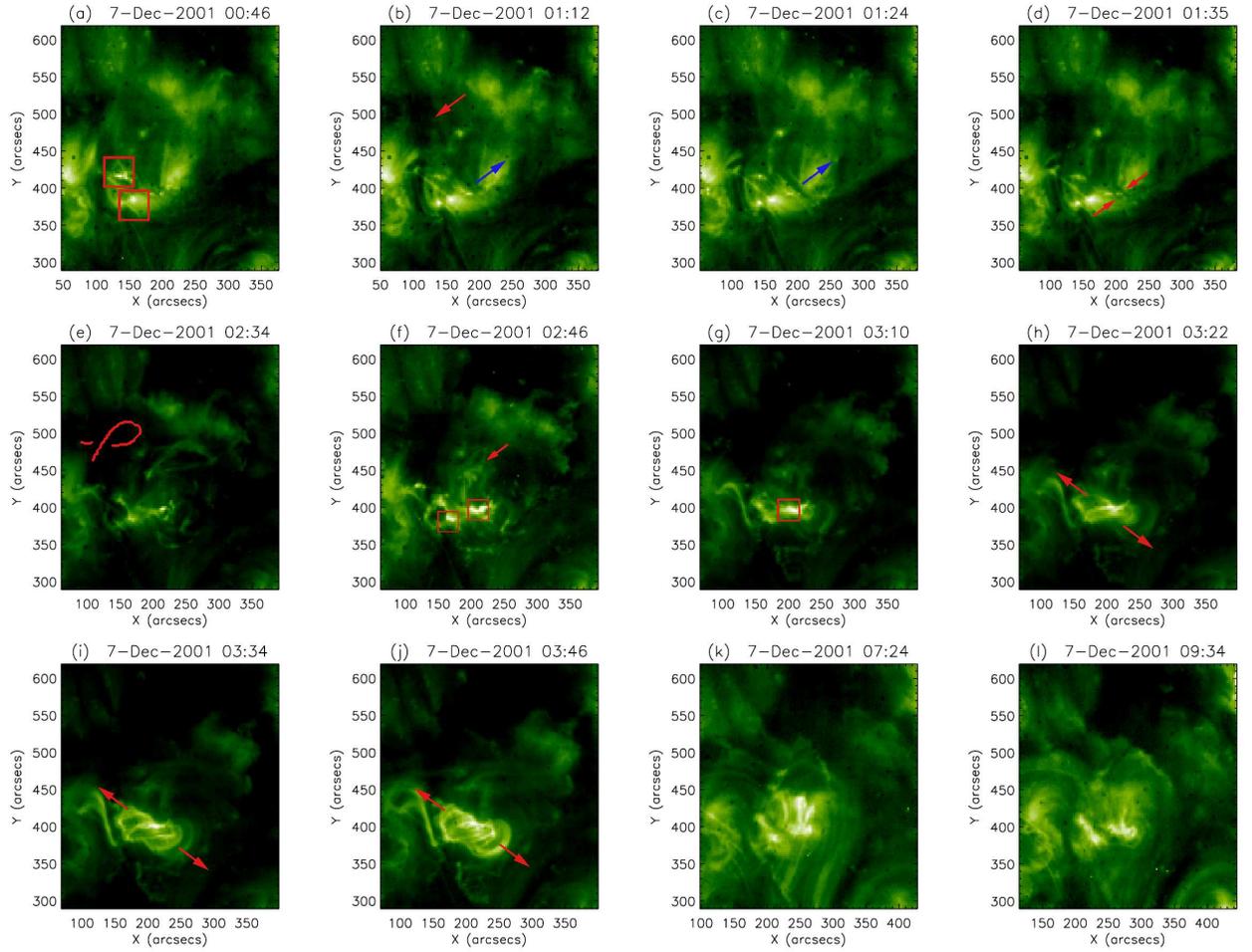}
\caption{EIT evolution in the eruption on December 07,2001.}
\label{fig:EIT}
\end{figure*}

\begin{figure*}[htb]
\centering
\includegraphics[angle=0,scale=1.0]{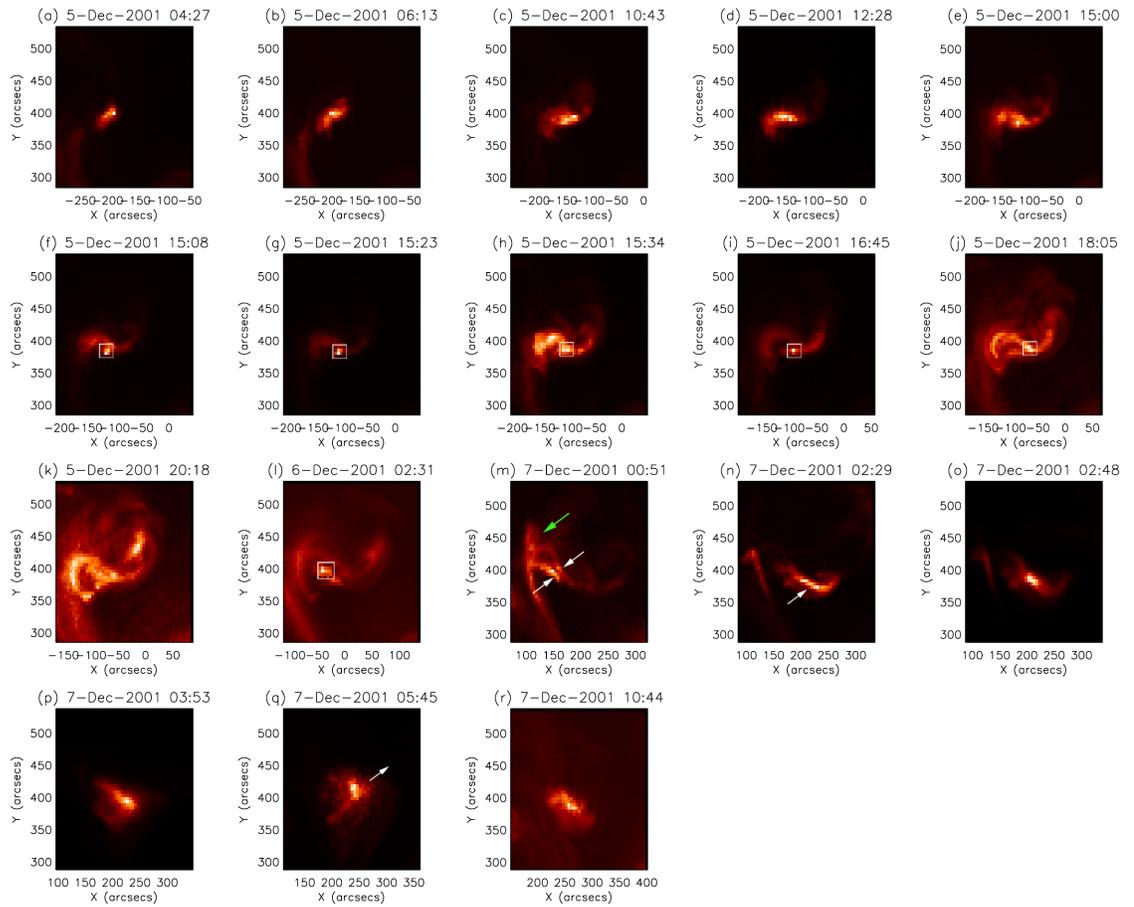}
\caption{Time Sequence of Yohkoh SXT images.} \label{fig:xrt}
\end{figure*}

 Fig.~\ref{fig:xrt} shows the evolution of NOAA 9729 in soft
xray. An Inverse S-shaped sigmoid structure formed clearly since
December 05. Such inverse S-shaped sigmoid is an indication of the
presence of negative twist in the magnetic field (Rust \& Kumar
1996; Pevtsov et al. 2001). The negative chirality is the same as
that deduced from the shape of filament in Fig.~\ref{fig:halpha}.
The shape of this sigmoid didn't change a lot in the following days
before eruption and it rotated clockwise which is consistent with
the rotation sense of connecting line of the two foot-points from WL
image as presented in Fig.~\ref{fig:mdi_WL}. An X-ray bright point
existed long time before eruption as labeled by the white square box
in Fig.~\ref{fig:xrt}f-i. At about 00:51UT on December 7, the left
part of the sigmoid risen up as denoted by the green arrow in Fig.
4m. The time is consistent with the eruption of left part of the
filament in the $\textrm{H}{\alpha}$ images Fig.~\ref{fig:halpha}g-j
and the EUV images Fig.~\ref{fig:EIT}b. At 02:29UT, the sigmoid has
erupted and a transient J-shaped sigmoid formed at the same time.
This transient J-shaped sigmoid existed earlier than the two bright
EUV foot-points denoted by the two red square boxes in
Fig.~\ref{fig:EIT}f and it became broad subsequently which can be
seen from Fig.~\ref{fig:EIT}o. At 03:53UT on December 07 in
Fig.~\ref{fig:EIT}p, a X-ray cusp has formed and it risen up firstly
as labeled by white arrow in Fig.~\ref{fig:EIT}q. This cusp was
diffused at last, which can be seen in Fig.~\ref{fig:EIT}r.
\subsection{CME evolution}
\label{sec:CME}
 Eruption of this active region brought one partial
halo CME which is recorded in CME catalog maintained at the CDAW
Data Center. The first appearance in the LASCO/C2 field of view
(FOV) is at 03:06UT on December 7 and the central position angle
(CPA) is $343^ \circ$. According to the description in this CME
catalog list, the linear speed obtained by fitting a straight line
to the height-time measurements is 803.2km/s. The acceleration of
such CME is -51.01$m/s^2$. This CME slows down within the LASCO FOV
while it has escaped out of ten solar radius. The eruptive part of
the filament escaped successfully. The speed value and deceleration
satisfy the character of the fast CMEs. Fast CMEs originate from an
active region and their initial speeds are well above the CME median
speed, 400 Km/s. They show no significant acceleration, but may show
some deceleration \citep{Cyr00}.

\subsection{Evolution of magnetic field and helicity}
\label{sec:helicity}

\begin{figure*}[htb]
\centering
\includegraphics[angle=0,scale=1.]{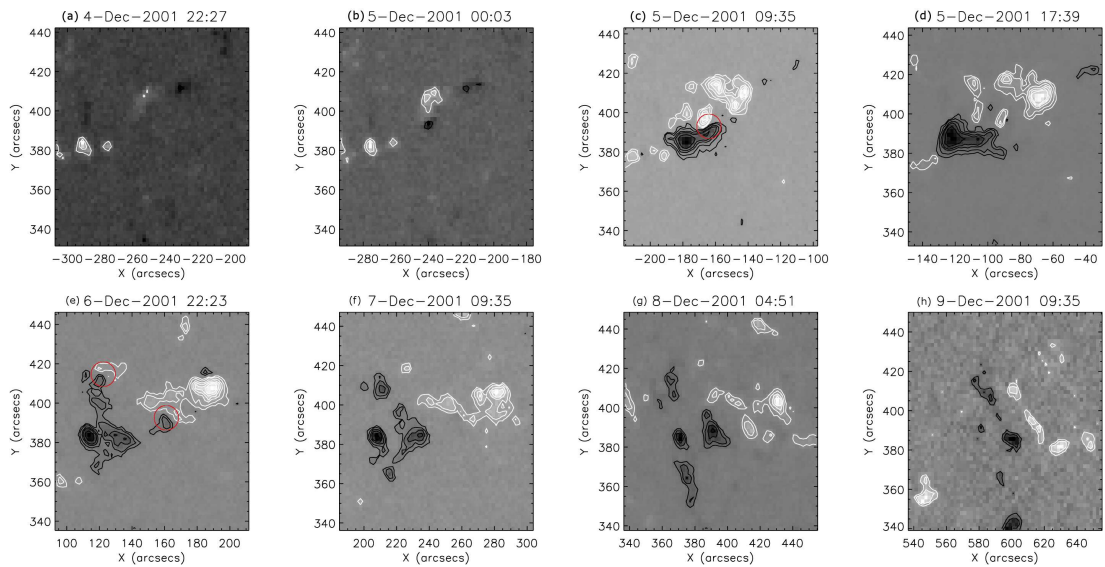}
\caption{Evolution of line-of-sight magnetic field from November 04
to 09, 2001. Positive (negative) magnetic flux is indicated by white
(black) solid lines at 100, 200, 400, 600 and 1000 G.}
\label{fig:mdi}
\end{figure*}

Fig.~\ref{fig:mdi} shows the evolution of line-of-sight magnetic
field of this active region. Two magnetic cancelation regions along
the PIL about four hours before flux tube eruption are labeled by
red circles in fig.3e. After the eruption the two cancelation
regions disappeared.

The accumulated magnetic helicity was calculated using full-disk
line-of-sight magnetograms taken by SOHO/MDI. From the photospheric
magnetic field observations the helicity flux across the photosphere
\textbf{S} can be calculated by
\begin{equation}
{\frac{dH_{R}}{dt}=-2\int(\vec{A}_{p}\cdot\vec{U}){B_n}\vec{dS},}
\label{eq:Hrate}
\end{equation}
where $\vec{U}$ denotes the horizontal velocity field. The vector
potential $\vec{A}_{p}$ is obtained by applying Local Correlation
Tracking (LCT) and Fast Fourier Transforms (FFT) to the normal
components of the photospheric magnetic field $B_n$ \citep{Cha01}.
After applying nonlinear mapping and flux density interpolation the
geometrical foreshortening was corrected \citep{Liu06,Yang09b}. To
reduce the noise, we set the horizontal velocity to zero in regions
where the magnetic field is small (${< 10 G}$). In order to better
track the emerging regions and to exclude the effect of relative
quiet regions outside the emergence sites, we set the horizontal
velocity to zero in regions of a weak cross-correlation (${<0.9}$)
of two magnetograms. We calculate the accumulated magnetic helicity
as
 \begin{equation}
 {H(t)=\int_{0}^{t}\frac{dH_{R}(t)}{dt}dt}
 \end{equation}
where the starting moment of time t=0 corresponds to the beginning
emergence of the active regions. Further, the following definition
of the normalized magnetic helicity $H_{norm}$ is used:
\begin{equation}
H_{norm}(t)  = \frac{{|H_m(t)|}}{{\Phi _m ^2 }}, \label{eq:Hnorm}
\end{equation}
where $\Phi_m$ is the maximum absolute value of the magnetic flux
through the photosphere of the studied newly emerging AR.

\begin{figure*}[htb]
\centering
\includegraphics[angle=0,scale=0.6]{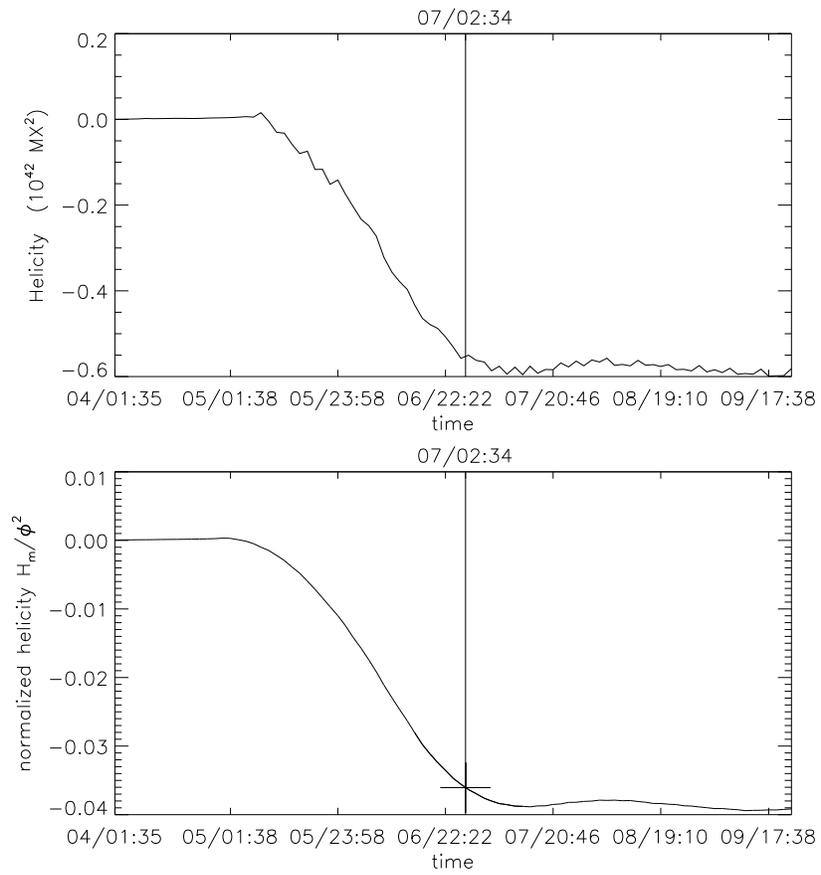}
\caption{The time profile of normalized parameter for NOAA 9729.}
\label{fig:hm_norm}
\end{figure*}

Fig.~\ref{fig:hm_norm} depicts the evolution of parameter that the
total relative magnetic helicity normalized by one-half of the
maximum sum of the unsigned positive and negative magnetic fluxes
$\phi$. This emerging flux tube take negative magnetic helicity to
the corona.  The flux tube in AR 9729 erupted at 02:34 UT on
December 07. Just at this moment, this normalized parameter is
-0.036. Recently, new methods have been used to calculate the
horizontal motion of magnetic structures on the photosphere, in
which the evolution of vertical magnetic fields satisfy the ideal
induction equation \citep[e.g.][]{Wel07}. The rotation motions of
the magnetic features, is introduced to get a more precise
measurement of the magnetic helicity \citep{Par05,Sch11,Rom11}. The
difference of magnetic helicity fluxes obtained from different
methods is usually within 15\% \citep{Rom11}. The estimate the final
normalized helicity is $-0.036\pm0.0054$ when the magnetic flux tube
erupted in AR 9729.
\section{SUMMARY AND DISCUSSION}
\label{sec:summary}

In this paper, we have presented a multiple wavelengths study of
emergence of a fast decayed  active region NOAA 9729 and associated
eruption. Our main observation results are summarized as follows.

1. NOAA 9729 emerged from convective zone with initial tilt angle is
almost perpendicular to the equator of sun. The connecting line
between leading polarity and following polarity rotated clockwise.
This active region still disobeyed Joy's law when it disappeared
from WL.

2. The filament in $\textrm{H}\alpha$, EUV structure and sigmoid in
X-ray all show the same inverse S-shape. This is a indication of
negative helicity in the flux tube. The eruption stages in the three
wavelength are also similar. The left part of the structure risen up
and the right part risen subsequently. Then, the eruption occurred
as the pre-eruption state evolved.

3. A filament in $\textrm{H}\alpha$ reformed again after erutpion.
The middle channel of filament in $\textrm{H}\alpha$ rotated
clockwise about $22^\circ$ than before.

4. EUV structure separated into two parts before eruption. The
sudden rising flux tube in the eruption showed a kinked structure.
The EUV bright regions disappeared after the flux tube risen up. The
risen flux tube dropped down and flare loops formed. The intimal
bright point formed above the PIL and expanded along the two
opposite directions of this PIL.

5. Inverse S-shaped sigmoid structure became thinner before
eruption. A transient J-shaped sigmoid structure formed subsequently
after eruption just before EUV bright region formed above the PIL
formed.

6. The associated CME recorded is a partial halo CME. The linear
speed obtained by fitting a straight line to the height-time
measurements is 803.2km/s. The acceleration of such CME is
-51.01$m/s^2$. Associated CME in the eruption belongs to the fast
CME.

7. From line-of-sight magnetic field, two magnetic cancelation
regions just underneath filament can be found before eruption and
these regions disappeared after eruption. Negative helicity was
taken by the emergence of magnetic flux and differential rotation.
The parameter that the total relative magnetic helicity normalized
by one-half of the maximum sum of the unsigned positive and negative
magnetic fluxes is -0.036 when eruption taken place.

What's the metastable structure before eruption?  Firstly, we can
deduced it as a negative twist flux tube. As described in the sum.
2, all structures in the emerging flux tube showed a inverse
S-shaped structure. These type of structures are associated with
negative twist magnetic flux tube. The negative accumulated magnetic
helicity also implies existence of negative twist in the flux tube.
Secondly, we can deduce this flux tube is a kinked flux tube. As the
description in the sum. 1, this active region disobeyed Joy's law
and the tilt angle rotated clockwise. If we consider this flux tube
as a simple flux tube described in \citet{Lop03}, such type of flux
tube should have a negative writhe (right hand) flux tube. In the
framework of a kink instability model as simulated by \citet{Fan99}
, also noted in \citet{Lek96} and \citet{Lin98}, the sign of the
writhe of kinked flux tubes would be the same as that of the twist
within the tubes due to conservation of helicity. When a horizontal
flux tube is emerging twisted right-handed (left-handed) of negative
(positive) magnetic helicity, the writhe of the tube axis resulting
from kink instability is also right-handed (left-handed). This would
leads to a clockwise (counter-clockwise) rotation of the apex
portion of the rising tube as viewed from the top. Such scenario
well explains the evolution in sum. 1. \citep{Yang09b} also pointed
out that the emerging flux tube tends to be caused by kink
instability if the accumulated helicity and writhe have the same
sign. Hence, we conclude that the metastable structure before
eruption is a kinked flux tube.

What's the trigger mechanism for this kinked flux tube? A kinked
flux tube can erupts because of kink instability. There are already
some evidences from observational results such as \citet{RL05} and
\citet{Liu07,Liu08} or from simulation results such as \citet{Fan05}
and \citet{TK05}. Another possible trigger mechanism is the torus
instability.  When the confining poloidal field decreases with
distance fast enough, radially outward perturbations of the flux
rope could trigger the torus instability \citep{KT06} before the
helical kink instability set in, and the toroidal flux rope would no
longer be confined. \citet{Fan07} performed 3D simulations to
investigate two distinct mechanisms that led to the eruption of the
flux tube. One case (case K) is the emerging flux rope is kinked and
a kink instability develops in it, leading to an eruption at last.
The other one (case T) is the overlying field declines more rapidly
with height, and the emerging flux rope is found to lose equilibrium
and erupt via the torus instability. The corresponding normal
relative magnetic helicity
 $H_m/\Phi ^2$ reaches approximately -0.16 for case
K and approximately -0.18 for case T. However, in our observation
result the normalized magnetic helicity is about -0.036 when the
flux tube in AR 9729 erupted, which is one order smaller than the
simulation results.

 \citet{Gil01} explored the various magnetic
configurations for failed and partial eruptions of filaments and
cavities. They suggested that reconnection at different positions of
the flux rope that threads into the filament creates different
topologies with implications of full, partial, or failed filament
eruptions. Reconnection occurs within the prominence can cause
partial filament to erupt and the rest part survive after eruption.
\citet{Moo01} also proposed a 3D Tether-Cutting model to explain the
eruption process of a sigmoid structure. They conjecture that the
magnetic explosion was unleashed by runaway tether-cutting via
implosive/explosive reconnection in the middle of the sigmoid, as in
the standard model. This internal reconnection apparently begins at
the very start of the sigmoid eruption and grows in step with the
explosion,caused the explosion to be ejective. In our observation,
the bright point in the middle of sigmoid structure could be found
clearly. Two magnetic filed cancelation regions also existed before
eruption and disappeared after eruption reflect the existence of
reconnection there. Therefore we propose that field lines rooted to
the photosphere near the inversion line where for the formation of a
magnetic tangential discontinuity are locally reconnection and cause
an instability. Field lines above the surface are detached from the
photosphere to form this CME and partially open the field which make
the filament loses equilibrium to rise quickly and then be drawn
back by the tension force of magnetic field after eruption to form a
new filament in our observation.

The eruption happened when the accumulated magnetic approach near
the maximum. It probably related to the possible existence of upper
bounds of total relative magnetic helicity for force-free magnetic
field in unbounded space, as conjectured in \citet{Zhang06} on the
study of axis-symmetric force-free field solutions. The absolute
normalized helicity is $0.036\pm0.054$, which is also close to the
theoretical prediction value 0.035 of \citet{Zhang08} when a
multipolar force-free magnetic field structure could sustain before
eruption.

\small
\section*{Acknowledgements}
{ This study is supported by grants 11078012,11173033,
11125314,10733020, 10921303, 41174153, 11103038 and 10673016 of
National Natural Science Foundation of China, and 2011CB811400 of
National Basic Research Program of China and $KLSA2010\_06$ of the
Collaborating Research Program of National Astronomical
Observatories, Chinese Academy of Sciences.}

\clearpage

\end{document}